\documentclass{elsart}

\usepackage{epsfig}
\usepackage{amssymb}
\newcommand{\bc}{\begin{center}}
\newcommand{\ec}{\end{center}}
\newcommand{\be}{\begin{equation}}
\newcommand{\ee}{\end{equation}}
\begin{document}

\begin{frontmatter}


 \title{Dijet cross sections in ep collisions:
who is afraid of symmetric cuts?
}
\author{Ji\v{r}\'{\i} Ch\'{y}la, Kamil Sedl\'{a}k
}
 \address{Institute of Physics, Academy of Sciences of the Czech Republic, \\
 18221 Na Slovance 2, Prague 8, Czech Republic
}




\begin{abstract}
Three widely used scenaria for defining cuts on transverse energies
of jets in ep collisions are discussed. All of them
are shown to suffer from the same sort of unphysical behaviour when
the cut regions are subject to additional constraints.
This feature is inherent in the very way dijet cross sections are defined
and cannot be avoided. In particular, the symmetric
cut scenario is shown to be equally suitable for the comparison with
NLO QCD calculations as the asymmetric or sum-like ones.
\end{abstract}

\end{frontmatter}

In analyzing the data on inclusive jet production in electron-proton
collisions some lower cut $E_c$ (at HERA typically $E_c\gtrsim 5$ GeV)
on the transverse energy of jets is necessary for experimental as well as
theoretical reasons. In parts of the multijet phase space the cut on the
transverse energies of two jets with highest $E_T$ may, however, get
into conflict with the infrared safety of the corresponding cross
sections.

In the next-to-leading QCD calculations of single and dijet cross
sections there are at most three final state partons, which combine to
two or three jets. Ordering the jets according to their transverse
energies in $\gamma^*$p center-of-mass system
($E_T^{(1)}\ge E_T^{(2)}\ge E_T^{(3)}$), the
corresponding phase space, marked by the upper
and lower solid straight lines in Figs. \ref{figsym}-\ref{figsum},
is given by the inequality $E_T^{(1)}/2\le E_T^{(2)}\le E_T^{(1)}$.
This constraint, given by pure kinematics, must then be combined
with additional conditions that might be imposed on jets for
experimental and
theoretical reasons. Phrasing the constraint on allowed part of
phase space in terms of  $E_T^{(1)},E_T^{(2)}$ does not, however,
imply that only genuine dijet variables must be investigated or only
the first two jets taken into account. For instance, the inclusive
single jet cross section gets in principle contribution even from the
third jet.

The aim of this letter is to argue that there is no way of choosing
the cuts on $E_T^{(1)},E_T^{(2)}$ that would be free from the sort of
unphysical behaviour of cross sections noted in \cite{Frixione}. We
shall illustrate our claims on the calculation of cross sections in
ep collisions at HERA in the region relevant for the analysis in
\cite{my}
\be
2\le
Q^2\le 80~{\mathrm{GeV}},~~0.1\le y\le 0.85,
~~
E_c=5~{\mathrm{GeV}},
\label{region}
\ee
but they are clearly of more general validity. All calculations
reported below were obtained by means DISENT NLO Monte-Carlo program
\cite{DISENT} using CTEQ6M set of parton distribution functions (PDF)
of the proton. For dijet events there are three classes of
distributions that can be measured:
\begin{itemize}
\item separate distributions of transverse energies and pseudorapidities
\be
\frac{{\mathrm{d}}\sigma}{{\mathrm{d}}E_T^{(1)}},~~
\frac{{\mathrm{d}}\sigma}{{\mathrm{d}}E_T^{(2)}},~~
\frac{{\mathrm{d}}\sigma}{{\mathrm{d}}\eta^{(1)}},~~
\frac{{\mathrm{d}}\sigma}{{\mathrm{d}}\eta^{(2)}},
\label{individual}
\ee
\item their sums, corresponding to distributions of what is
called in \cite{Kramer} ``trigger jets'',
\be
\frac{{\mathrm{d}}\sigma}{{\mathrm{d}}E_T}\equiv
\frac{{\mathrm{d}}\sigma}{{\mathrm{d}}E_T^{(1)}}+
\frac{{\mathrm{d}}\sigma}{{\mathrm{d}}E_T^{(2)}},~~
\frac{{\mathrm{d}}\sigma}{{\mathrm{d}}\eta}\equiv
\frac{{\mathrm{d}}\sigma}{{\mathrm{d}}\eta^{(1)}}+
\frac{{\mathrm{d}}\sigma}{{\mathrm{d}}\eta^{(2)}},
\label{both}
\ee
\item or the distributions of some of the combinations characterizing
the whole event, like their mean values
$\overline{E}_T\equiv (E_T^{(1)}+E_T^{(2)})/2$
and $\overline{\eta}\equiv (\eta^{(1)}+\eta^{(2)})/2$
\be
\frac{{\mathrm{d}}\sigma}{{\mathrm{d}}\overline{E}_T},~~
\frac{{\mathrm{d}}\sigma}{{\mathrm{d}}\overline{\eta}},
\label{means}
\ee
but one is free to use general combinations
as well.
\end{itemize}

\begin{figure}\centering\unitlength=1mm
\begin{picture}(140,70)
\put(-7,0){\epsfig{file=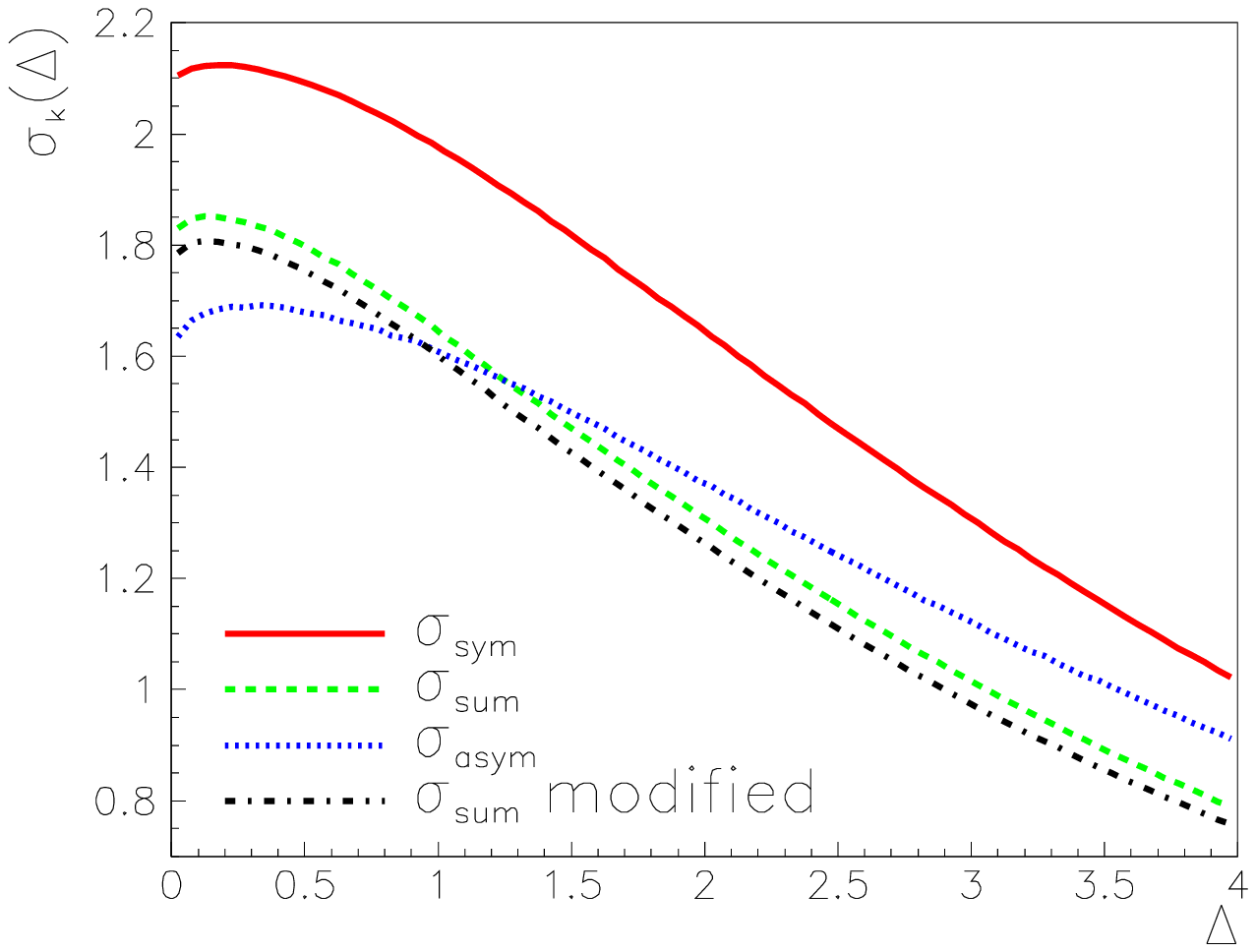,width=9.6cm}}
\put(93,0){\epsfig{file=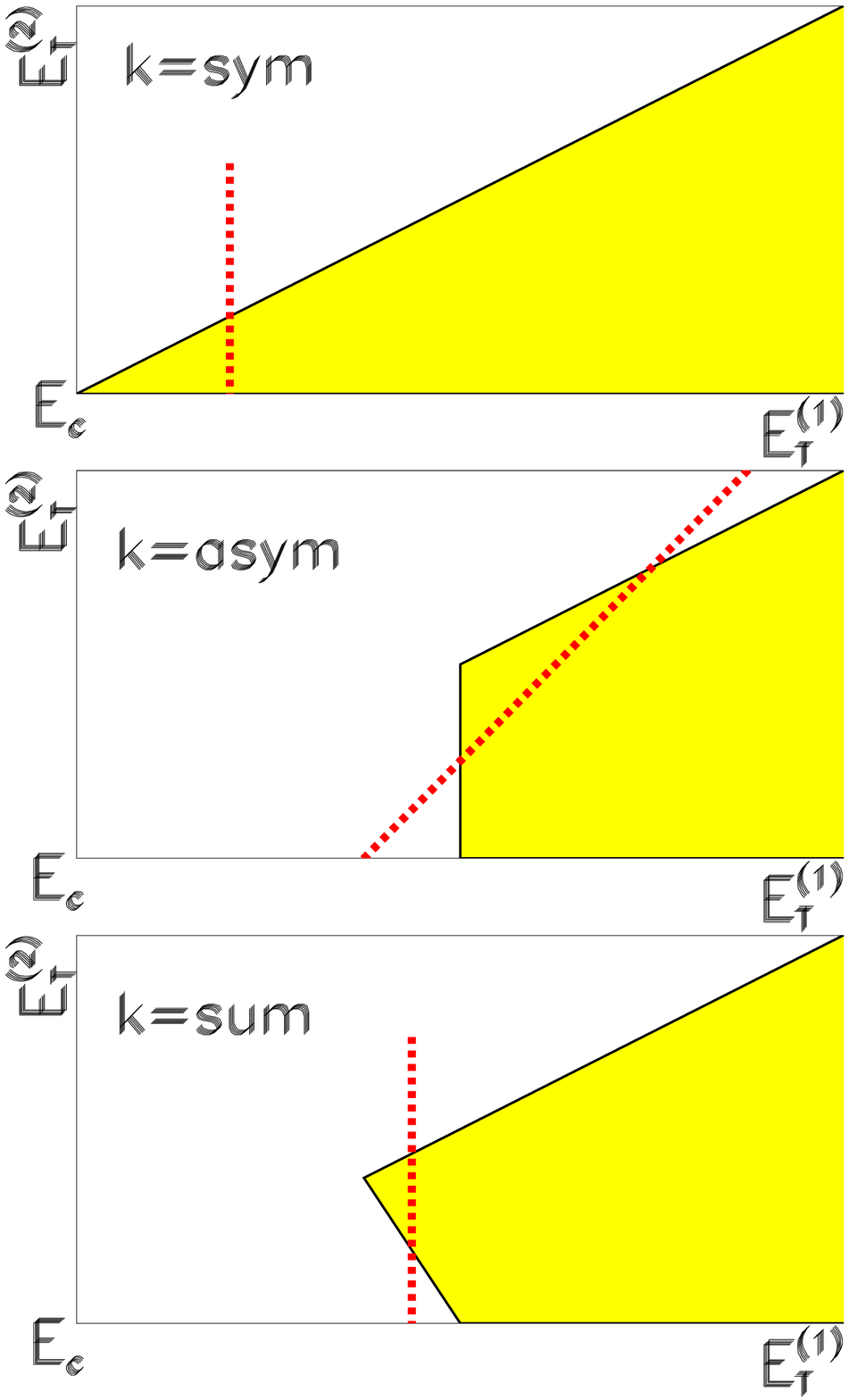,width=4.7cm}}
\end{picture}
\caption{The dependence of $\sigma_k(\Delta)$ on the parameter
$\Delta$ restricting the regions defined in
(\ref{sym},\ref{asym},\ref{sum}) and marked in yellow, to the right
of the red dotted curves.}
\label{fig1}
\end{figure}

\begin{figure}[h]\centering \unitlength=1mm
\begin{picture}(130,60)
\put(0,0){\epsfig{file=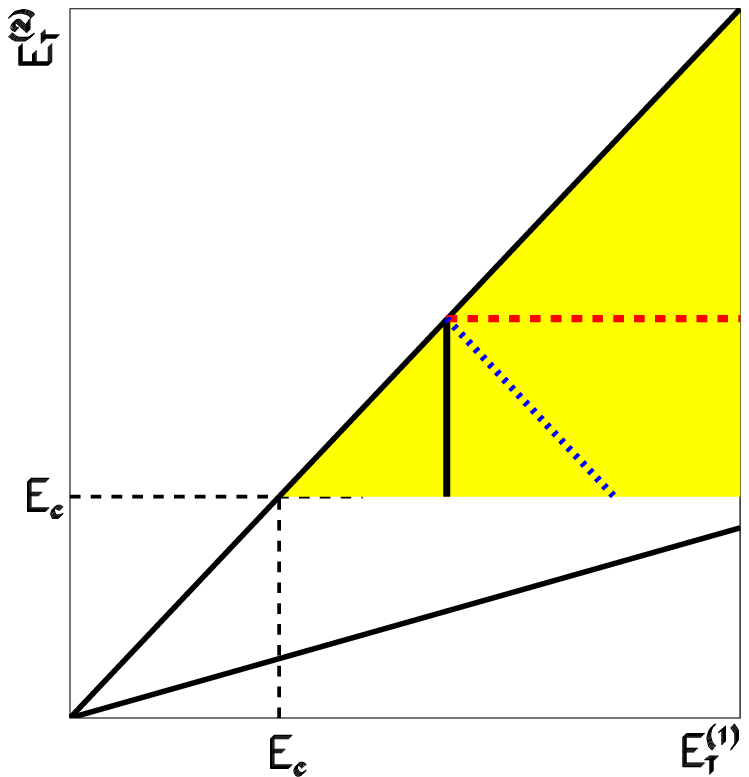,width=6cm}}
\put(65,0){\epsfig{file=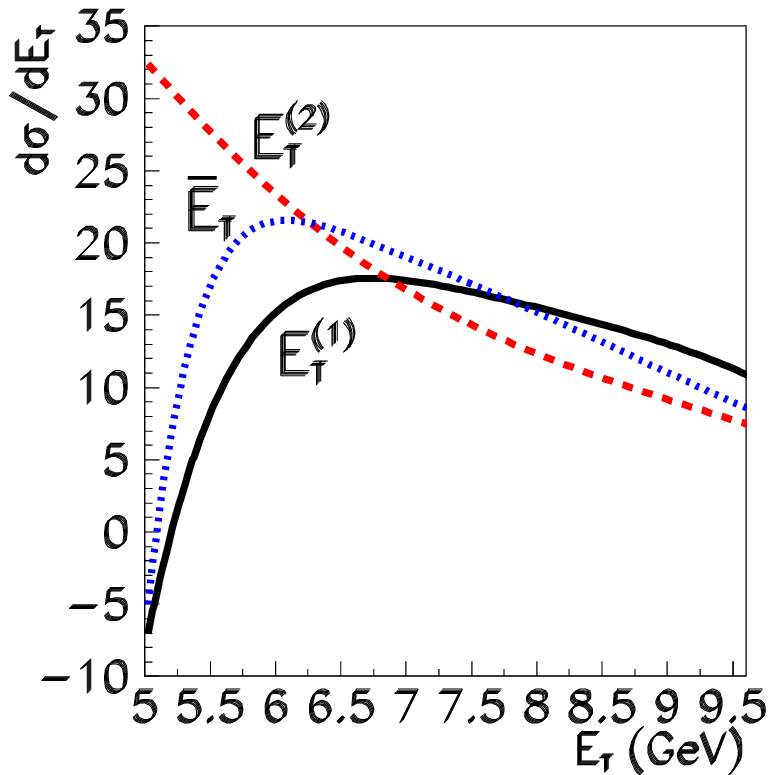,width=6cm}}
\end{picture}
\caption{Left: symmetric cuts scenario, $E_c=5$ GeV. The solid, dotted
and dashed lines correspond to fixed values of
$E_T^{(1)},\overline{E}_T$ and $E_T^{(2)}$. As they approach $E_c$
from above the lengths of the former two decrease, whereas that of
$E_T^{(2)}$ increases. Right: corresponding distributions of
$E_T^{(1)}$ (solid), $E_T^{(2)}$ (dashed) and $\overline{E}_T$ (dotted).}
\label{figsym}
\end{figure}

\begin{figure}\centering \unitlength=1mm
\begin{picture}(120,60)
\put(0,0){\epsfig{file=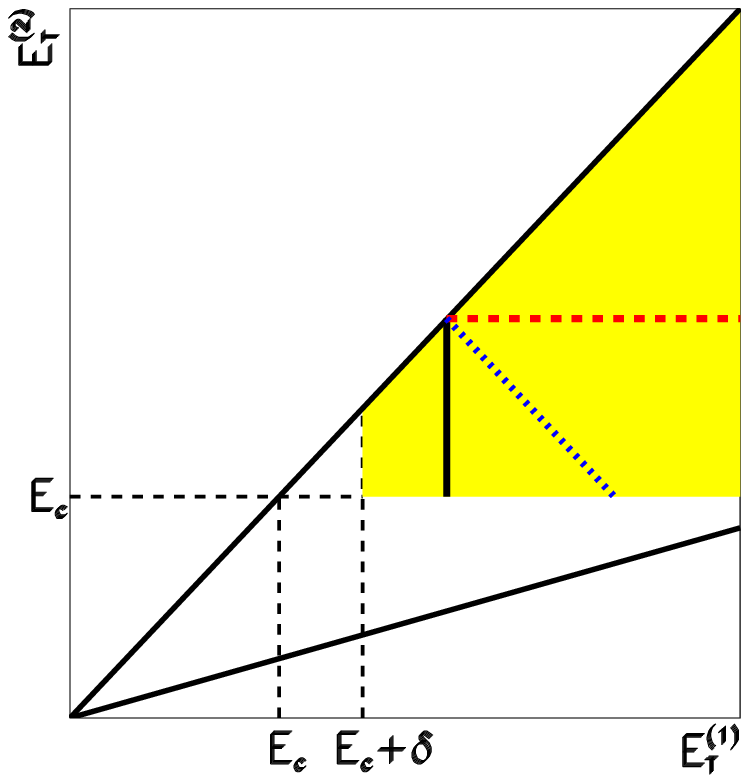,width=6cm}}
\put(65,0){\epsfig{file=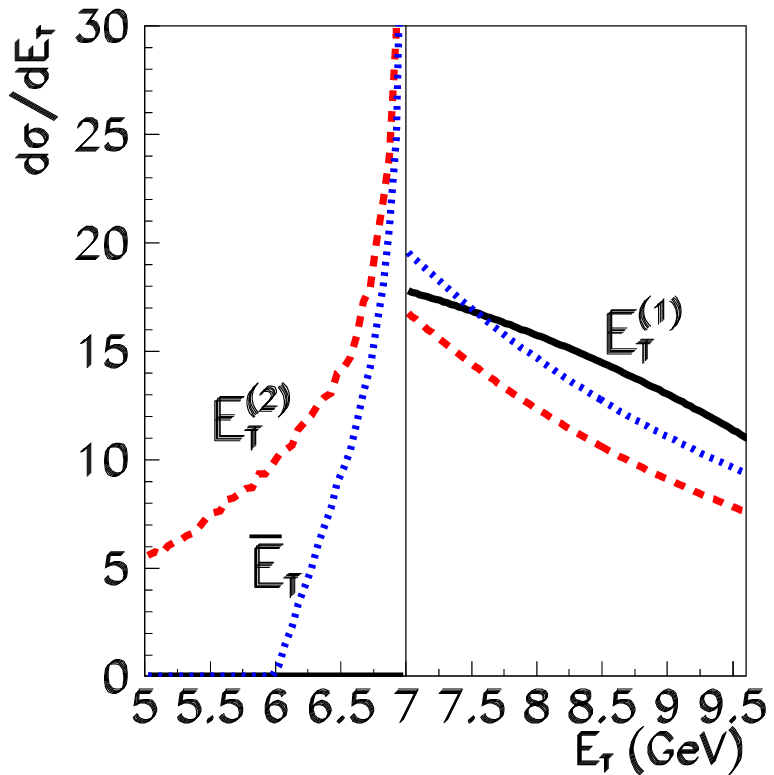,width=6cm}}
\end{picture}
\caption{The same as in Fig. \ref{figsym} but for asymmetric cuts
scenario with $E_c=5$ GeV and $\delta=2$ GeV.}
\label{figasym}
\end{figure}

\begin{figure}\centering \unitlength=1mm
\begin{picture}(120,60)
\put(0,0){\epsfig{file=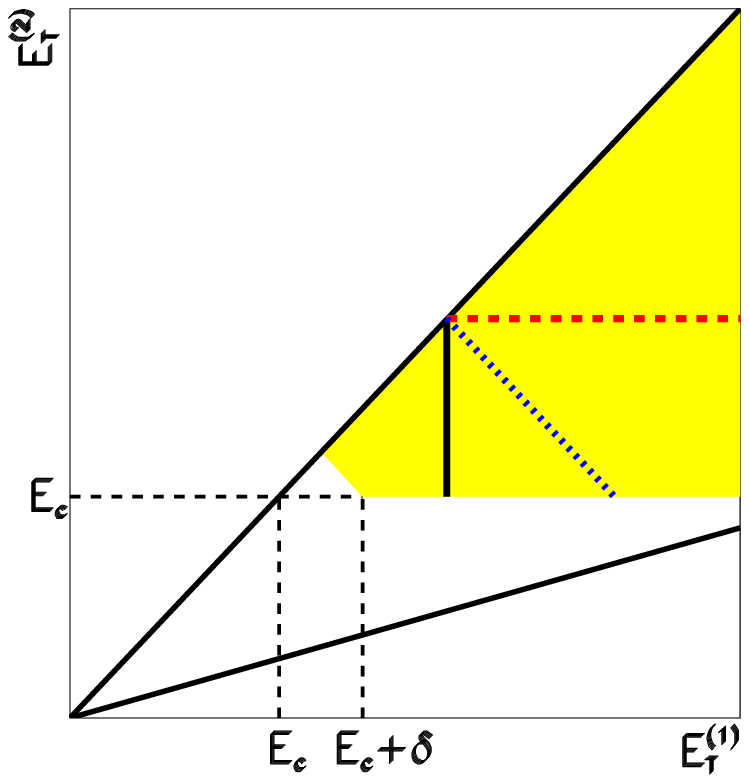,width=6cm}}
\put(65,0){\epsfig{file=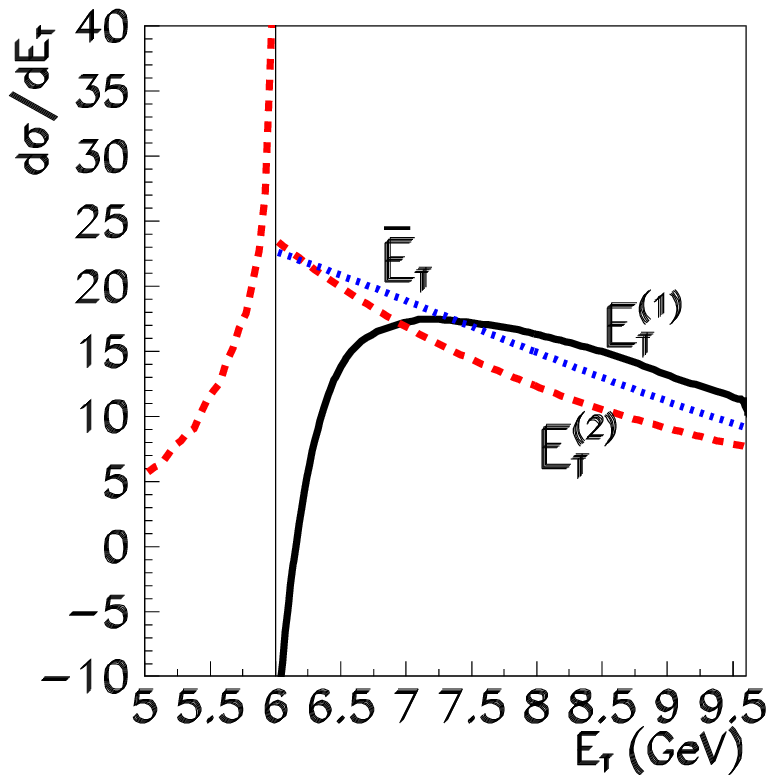,width=6cm}}
\end{picture}
\caption{The same as in Fig. \ref{figsym} but for the sum cuts
scenario with $E_c=5$ GeV and $\delta=2$ GeV.}
\label{figsum}
\end{figure}

\bc {\em Symmetric cut scenario}\ec
In this scenario the same cut is imposed on both jets
\be
E_T^{(1)}\ge E_c,~~~E_T^{(2)}\ge E_c.
\label{sym}
\ee
This is quite appropriate for the comparison of data with lowest
QCD calculations because the latter involve binary partonic hard processes
which lead (in diparton rest frame) to two partons
with the same transverse energies. Its use in NLO QCD calculations of
dijet cross sections has been questioned in \cite{Frixione} observing
that it leads to unphysical behaviour of the cross section
$\sigma_{sym}(\Delta)$ corresponding to the integral over the region
defined in (\ref{sym}) supplemented with an additional constraint
$E_T^{(1)}\ge E_c+\Delta$, represented graphically by
the red dotted curve in the upper right plot of Fig. \ref{fig1}.
The resulting dependence, displayed as red solid curve in the left part
of Fig. \ref{fig1}, implies that for $\Delta$ approaching
zero, $\sigma_{sym}(\Delta)$ decreases despite the fact that the
corresponding phase space increases.
The source of this unphysical behaviour can be traced back to the fact
that at the order $\alpha\alpha_s^2$ perturbative QCD does not
yield a well-defined result for the double differential cross section
${\mathrm{d}}\sigma/{\mathrm{d}}E_T^{(1)}{\mathrm{d}}E_T^{(2)}$ when
$E_T^{(1)}=E_T^{(2)}$. This comes from the fact that at this order
there are mass divergencies of the virtual corrections that
must be cancelled by
those of the real emissions. For a quantity to be calculable in
perturbative QCD one must thus allow for integration over sufficient
part of the three parton phase space, where $E_T^{(2)}< E_T^{(1)}$.

To understand better the origin of the behaviour of $\sigma_{sym}(\Delta)$
we plot in Fig.~\ref{figsym} the distributions of
$E_T^{(1)},E_T^{(2)}$ and $\overline{E}_T$ for $E_c=5$~GeV. As
expected, the distributions of both $E_T^{(1)}$ and $\overline{E}_T$ turn
negative for $E_T^{(1)}$ or $\overline{E}_T$ close to $E_c=5$ GeV because of
the small volume (represented in left parts of Figs. \ref{figsym}-\ref{figsum}
by the length of perpendicular solid or oblique dotted lines)
of phase space available for the real parton emission. In
fact both these distributions diverge to $-\infty$ as $E_T^{(1)}$ or
$\overline{E}_T$ approach $E_c$ from above, thereby causing
$\sigma_{sym}(\Delta)$ and $\sigma_{sum}(\Delta)$ in Fig. \ref{fig1}
to bend down for small $\Delta$. On the other hand
${\mathrm{d}}\sigma/{\mathrm{d}}E_T^{(2)}$ grows monotonously with
decreasing $\Delta$, reflecting the fact that even for $E_T^{(2)}=E_c$ there
is plenty of phase space (represented in the left
parts of Figs. \ref{figsym}-\ref{figsum} by the dash horizontal lines)
available for the real gluon emission because $E_T^{(1)}$ can be
anywhere above it. Note that the integrals over all three distributions
in Fig. \ref{figsym} are finite and equal as they correspond to
three different ways of performing the integral over the same region defined
in (\ref{sym}). There is thus no reason to reject the symmetric cut scenario
for the comparison of data with NLO QCD, but for $E_T$ close to $E_c$,
one should
compare the distributions of $E_T^{(2)}$, rather than those of $E_T^{(1)}$.

\bc{\em Asymmetric cut scenario}\ec
To avoid the above unphysical behaviour of $\sigma_{sym}$ the asymmetric cut
scenario
\be
E_T^{(1)}\ge E_c+\delta,~~~E_T^{(2)}\ge E_c,
\label{asym}
\ee
where $\delta\ge 0$, was suggested in \cite{Frixione}
\footnote{Instead of (\ref{asym}) slightly different modification of
(\ref{sym}) was used even earlier in \cite{Kramer}.}.
In Fig. \ref{figasym} we plot the corresponding results for the same
three distributions as in Fig. \ref{figsym}. As expected, the
distribution of $E_T^{(1)}$ now behaves physically down to $E_c+\delta$,
as do those of $E_T^{(2)}$ and $\overline{E}_T$. However, whereas the
first distribution ends there, the other two continue below $E_c$. Both
of them are, however, singular at
$E_T^{(2)}=\overline{E}_T=E_c$, diverging at this point from left to
$+\infty$ due to the divergence of real emission contributions as
$E_T^{(2)}\rightarrow E_T^{(1)}$. However,
as in the case of the symmetric cuts, the integrals over the
distributions in Fig. \ref{figasym} are finite and the same.
The behaviour of the distributions of
$E_T^{(2)}$ and $\overline{E}_T$ below $E_c$ is of course as unphysical
as the negative values of the distributions of  $E_T^{(1)}$ and
$\overline{E}_T$ closely above $E_c$ in the
symmetric cut scenario of Fig. \ref{figsym}.

Moreover, it is simple to invent an additional constraint that cuts
off part of the phase space region defined in (\ref{asym}) in a similar
way as the condition $E_T^{(1)}\ge E_c+\Delta$ in the case of the
symmetric cut. For instance the restriction
\be
E_T^{(2)}\le 2E_T^{(1)}-E_c-\delta-\Delta,
\label{addasym}
\ee
which cuts off part of the region (\ref{asym}) as shown in the middle
right picture in Fig. \ref{fig1} results in the
integrated cross section $\sigma_{asym}(\Delta)$ exhibiting the same
kind of unphysical behaviour for $\Delta\rightarrow 0$ as
$\sigma_{sym}(\Delta)$.

\bc{\em Sum cut scenario}\ec
The cuts on $E_T^{(1)},E_T^{(2)}$ can also be formulated in terms of
their mean value
\be
\overline{E}_T\equiv \frac{E_T^{(1)}+E_T^{(2)}}{2}
\ge E_c+\frac{\delta}{2},~~~E_T^{(2)}\ge E_c.
\label{sum}
\ee
As shown in Fig. \ref{figsum}
in this case it is the distributions in $E_T^{(1)}$ and $E_T^{(2)}$
that are singular at $E_c$: the
former turns negative and diverges when $E_T^{(1)}$ approaches
$E_c$ from above, whereas the latter diverges to plus infinity
for $E_T^{(2)}\rightarrow E_c$ from below. As in the previous two
cases the integrals over all three distributions are, however,
finite and the same.

Although even for the lowest $\overline{E}_T$ the region defined
in (\ref{sum}) involves integration over finite neighbourhood of
the point $E_T^{(1)}=E_T^{(2)}$, this scenario was claimed in
\cite{Poetter} not to be infrared safe. The argument was based on
the fact that similarly as in the case of the symmetric cut
scenario (\ref{sym}) the integrated cross section
$\sigma_{sum}(\Delta)$, corresponding to the additional constraint
$E_T^{(1)}\ge E_c+\delta/2+\Delta$ behaves, as illustrated
by the green dashed curve in Fig. 1, unphysically for
$\Delta\rightarrow 0$. However, as emphasized above, the same
objection can actually be raised against any scenario. This is clear
from the geometrical meaning of the additional cuts involving the
parameter $\Delta$: as shown in the right part of Fig. \ref{fig1}, in
all three scenaria it cuts off the peaks of the wedges in the plane
$E_T^{(1)},E_T^{(2)}$ where the negative infinities of the virtual
corrections dominate.

The scenario (\ref{sum}) is often supplemented with a
constraint on the relative difference of jet transverse energies,
for instance, $(E_T^{(1)}-E_T^{(2)})/(E_T^{(1)}+E_T^{(2)})\le 1/4$.
This cuts off part of the three jet phase space but as
indicated by the black dash-dotted curve in Fig. \ref{fig1},
has only small effect of the behaviour of $\sigma_{sum}(\Delta)$.

The results presented in Figs. \ref{fig1}-\ref{figsum} demonstrate
that there is no real advantage of the asymmetric or sum-like cut
scenaria over the symmetric one. All of them suffer from the same
kind of unphysical behaviour of two of the three distributions in
$E_T^{(1)},E_T^{(2)}$ or $\overline{E}_T$ and all three can be
supplemented by additional constraints that lead to the same
unphysical dependence on integrated cross sections $\sigma_{sym}(\Delta)$,
$\sigma_{asym}(\Delta)$, or $\sigma_{sum}(\Delta)$ displayed in Fig.
\ref{fig1}.

\vspace*{0.5cm}
\bc{\em Distributions in bins of finite width}\ec
In practice experimental distributions in $E_T^{(1)},E_T^{(2)}$
and $\overline{E}_T$ have finite bin widths. For each bin of these
variables we can introduce an additional cut $\Delta$, which chops off
part of this bin, and investigate the dependence of the bin content on
this parameter in the limit $\Delta\rightarrow 0$, much in the same
way as we did above for the integrated cross sections. Note that for
$E_T^{(1)},E_T^{(2)}\ge E_c+\delta$ the choice of the cut scenario is
irrelevant for these considerations as it does not influence the
distributions there. Let us take as an example the bin
\begin{figure}\centering \unitlength=1mm
\begin{picture}(120,50)
\put(-13,0){\epsfig{file=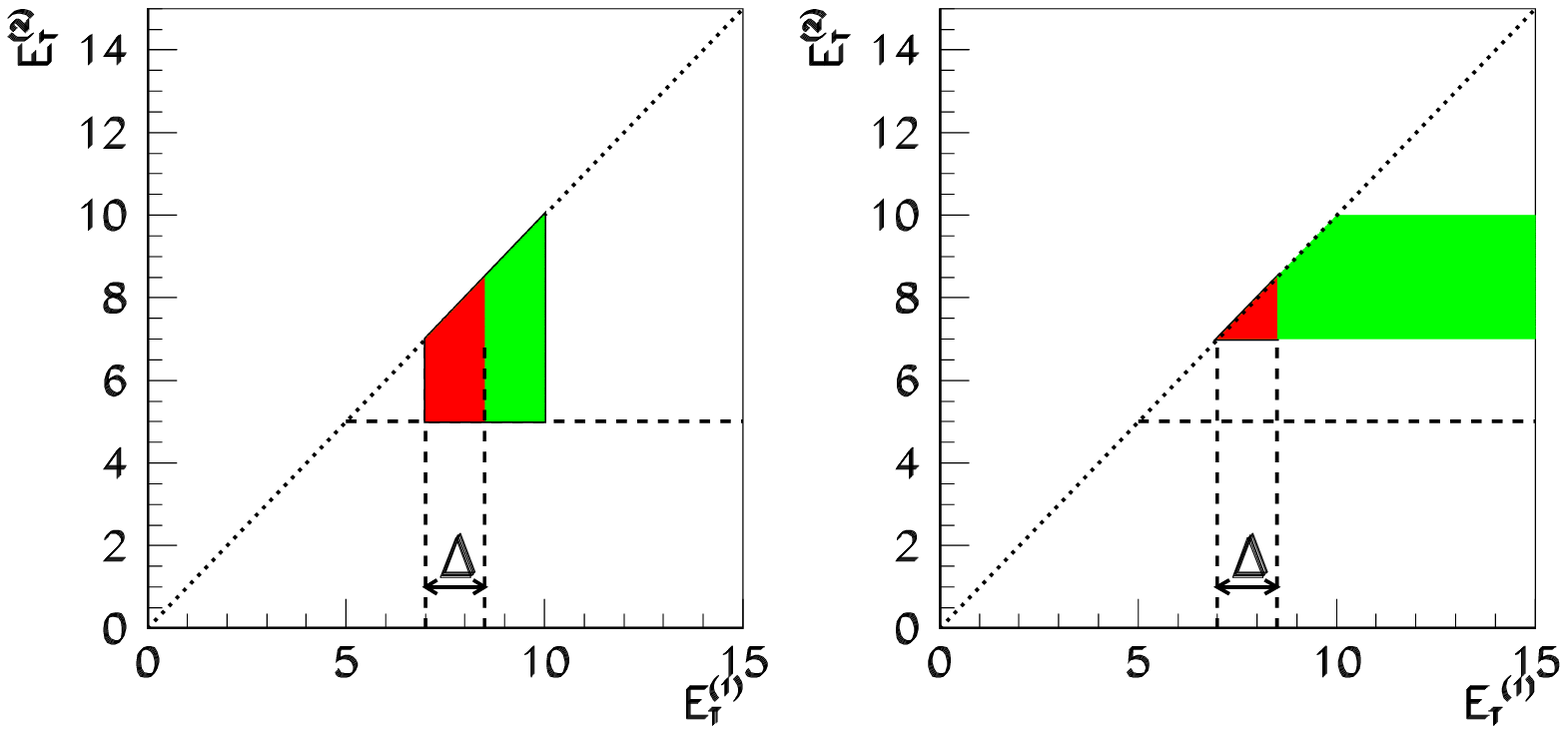,width=10cm}}
\put(87,5){\epsfig{file=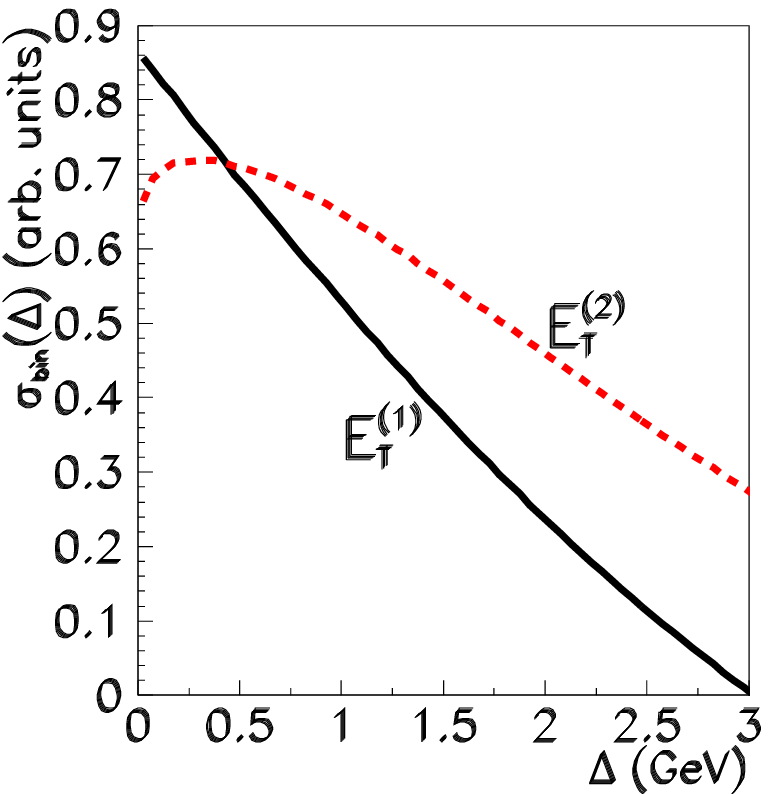,width=4.5cm}}
\end{picture}
\caption{The effects of the cutting the bin defined in
(\ref{binn}-\ref{bin}) according to (\ref{bin1}) on the
distributions in $E_T^{(1)}$ and $E_T^{(2)}$.}
\label{plots1}
\end{figure}
\begin{eqnarray}
7 \le E_T^{(1)}\le 10~{\mathrm{GeV}},&~~~&E_T^{(2)}\ge E_c
\label{binn}\\
7 \le E_T^{(2)}\le 10~ {\mathrm{GeV}},&~~~&E_T^{(1)}\ge E_T^{(2)}
\label{bin}
\end{eqnarray}
represented in left two pictures of Figs. \ref{plots1},\ref{plots2}
by the vertical and horizontal
 colored bands. We first introduce the
parameter $\Delta$ similarly as for the symmetric cut (\ref{asym}) by
imposing additional constraint:
\begin{equation}
7+\Delta \le E_T^{(1)}\le 10~ {\mathrm{GeV}}
\label{bin1}
\end{equation}
which means graphically cutting off the red areas in Fig. \ref{plots1}.
The resulting $\Delta$-dependence of the contributions
${\mathrm{d}}\sigma/dE_T^{(1)}$ and ${\mathrm{d}}\sigma/dE_T^{(2)}$
to the corresponding bin content $\sigma_{bin}(\Delta)$,
displayed in the right part of Fig. \ref{plots1}, shows that the
distribution of $E_T^{(2)}$ behaves unphysically for small
$\Delta$, whereas that of $E_T^{(1)}$ increases monotonously as
$\Delta\rightarrow 0$.

However, we can cut off part of the bin
(\ref{binn}-\ref{bin}) in an alternative way represented by the
red areas in Fig. \ref{plots2}
\begin{equation}
7 \le E_T^{(2)}\le 10-\Delta~ {\mathrm{GeV}},
\label{bin2}
\end{equation}
which is certainly as legitimate as the cut (\ref{bin1}).
The corresponding
results, shown in the right part of Fig. \ref{plots2}, lead, however,
to opposite conclusion! Now it is the
${\mathrm{d}}\sigma/dE_T^{(2)}(\Delta)$ distribution that is a
monotonously decreasing function of $\Delta$, whereas
${\mathrm{d}}\sigma/dE_T^{(1)}(\Delta)$ exhibits much
the same unphysical dependence on $\Delta$ as did
${\mathrm{d}}\sigma/dE_T^{(2)}(\Delta)$ for the cut defined in (\ref{bin1}).
\begin{figure}[h]\centering \unitlength=1mm
\begin{picture}(120,50)
\put(-13,0){\epsfig{file=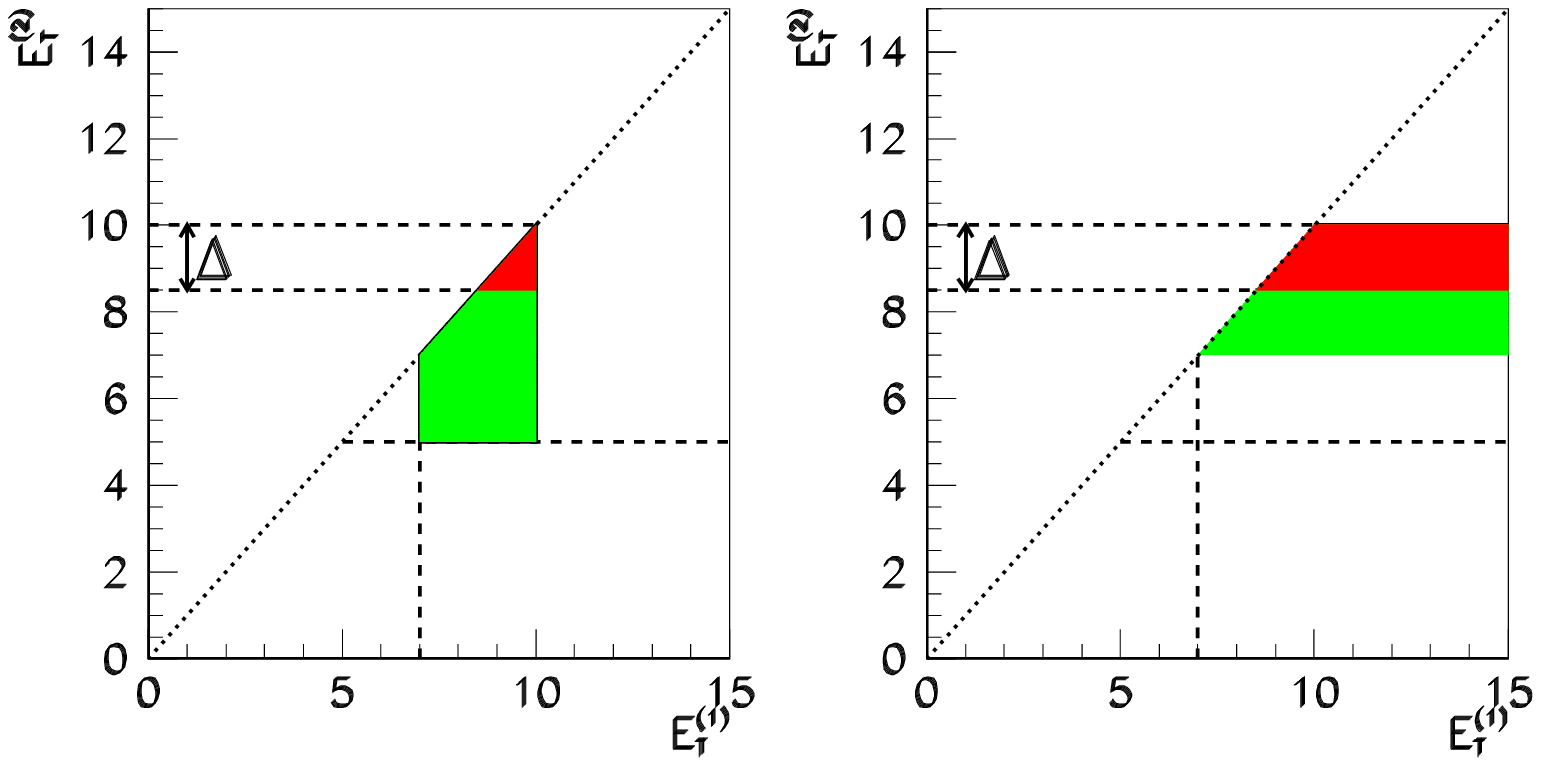,width=10cm}}
\put(87,3){\epsfig{file=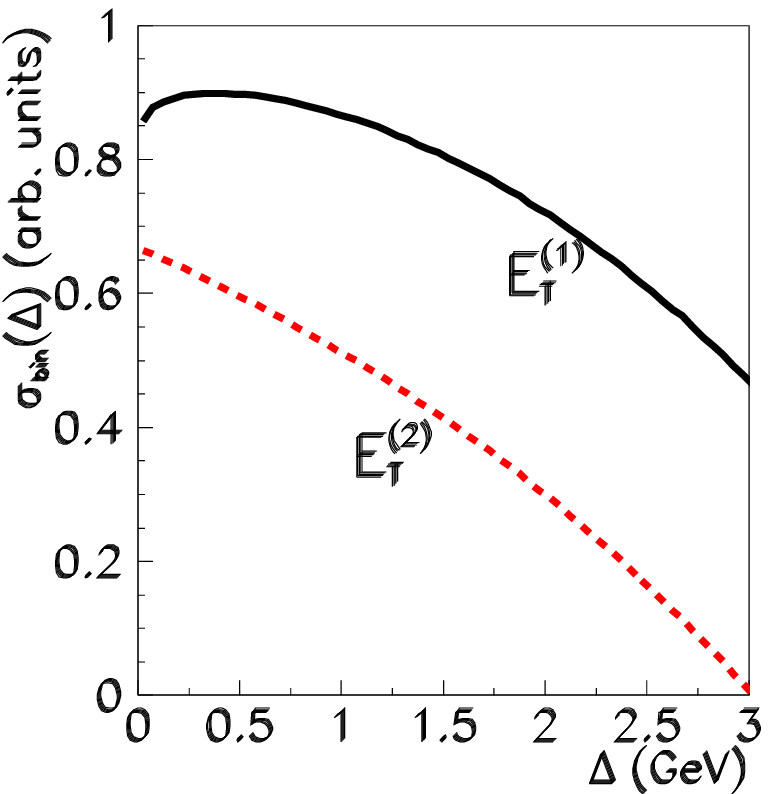,width=4.6cm}}
\end{picture}
\caption{
The same as in Fig. \ref{plots1} but for the cut defined in
(\ref{bin2}).}
\label{plots2}
\end{figure}

In summary, for dijet sample there is thus no way how to introduce the cuts
on the jet transverse energies that would avoid entirely the problem noted in
\cite{Frixione}. Whatever variables, quantity and binning we choose, there
is always a simple way how to cut the chosen kinematical region in a way that
leads to the same sort of unphysical behaviour of cross sections shown in
Fig. \ref{fig1} and the asymmetric cut scenario is therefore not superior to
the symmetric one. On the
other hand the above exercise also suggest a simple cure: one just needs to
choose the appropriate variable or take sufficiently wide bins in jet
transverse energies.

We are grateful to G. Grindhammer and V. Shekelyan for raising the
questions that initiated
this investigation. This work was supported in part by the project
LN00A006 of the Ministry of Education of the Czech Republic and by the
Institutional research project AV0Z1-010-920.


\begin{thebibliography}{00}
\bibitem{Frixione} S. Frixione, G. Ridolfi, Nucl. Phys. B 507 (1997) 315.
\bibitem{my} C. Adloff et al. (H1 Collaboration), in preparation
\bibitem{DISENT} S. Catani, M. H. Seymour, Nucl. Phys. B 485 (1997) 291.
\bibitem{Kramer} M. Klasen, G. Kramer, Phys. Lett. B 366 (1996) 385.
\bibitem{Poetter} B.P\"{o}tter, Comp. Phys. Commun. 133 (2000) 105.
\end{thebibliography}
\end{document}